\documentclass{aastex}          
\usepackage{spr-astr-addons}    

\usepackage[T1]{fontenc}
\usepackage{ae,aecompl}
\usepackage{amsmath}

\usepackage{graphicx}	
\usepackage{amsmath}	
\usepackage{amssymb}	
\usepackage{array}
\usepackage{tabularx}

\defcitealias{2011ApJ...733L..17L}{LCK11}
\defcitealias{2014MNRAS.445.4218K}{K14}
\defcitealias{2014ApJ...791L..39J}{J14}

\newcommand{\emaila}{}

\begin{document}

\title{The X-ray bursts within the 2010 outburst of the accreting millisecond X-ray pulsar SAX J1748.9-2021}

\shorttitle{X-ray bursts from the AMXP SAX J1748.9-2021}
\shortauthors{Wu et al.}

\author{Ziwei Wu\altaffilmark{$^\dagger$,1,2}}
\and
\author{Guoqiang Ding\altaffilmark{1}}
 \and
\author{Zhibing Li\altaffilmark{1}}
\and
\author{Yupeng Chen\altaffilmark{3}}
\and
\author{Jinlu Qu\altaffilmark{3}}

\altaffiltext{$^\dagger$}{\emaila{E-mail: wuziwei@xao.ac.cn}}
\altaffiltext{1}{Xinjiang Astronomical Observatory, Chinese Academy of Sciences,150, Science 1-Street, Urumqi, Xinjiang 830011, China}
\altaffiltext{2}{University of Chinese Academy of Sciences, Beijing 100049, China}
\altaffiltext{3}{Key Laboratory for Particle Astrophysics, Institute of High Energy Physics, Chinese Academy of Sciences, Beijing 100049, China}

\begin{abstract}
With the observations from \textit{Rossi X-ray Timing Explorer}, we search and study the X-ray bursts of accreting millisecond X-ray pulsar SAX~J1748.9-2021 during its 2010 outburst. We find 13 X-ray bursts, including 12 standard type-\uppercase\expandafter{\romannumeral1} X-ray bursts and an irregular X-ray burst which lacks cooling tail. During the outburst, the persistent emission occurred at $\sim$(1-5)$\%\rm {\dot{M}_{Edd}}$. We use a combination model of a blackbody (BB), a powerlaw, and a line component to fit the persistent emission spectra. Another BB is added into the combination model to account for the emission of the X-ray bursts due to the thermonuclear burning on the surface of the neutron star. Finally, we modify the combination model with a multiplicative factor $f_{\rm a}$, plus a BB to fit the spectra during the X-ray bursts. It is found that the $f_{\rm a}$ is inversely correlated with the burst flux in some cases. Our analysis suggests that the ignition depth of the irregular X-ray burst is obviously smaller than those of the type-\uppercase\expandafter{\romannumeral1} X-ray bursts. We argue that the detected type-\uppercase\expandafter{\romannumeral1} X-ray bursts originate from helium-rich or pure-helium environment, while the irregular X-ray burst originates from the thermonuclear flash in a shallow ocean.
\end{abstract}




\section{Introduction}
\label{intro}
Neutron star (NS) low mass X-ray binaries (NS-LMXBs) consist of a magnetized NS ($\mathrm{B_{NS}<10^{10}}$ G) and a low mass companion star ($\mathrm{M_{c}\,\leq\,1M_{\odot}}$). Soft-to-hard transitions are commonly seen in NS and black hole X-ray binaries \cite[e.g.,][]{2007ApJ...667.1073L,2001ApJS..132..377H}. Accreting millisecond X-ray pulsars (AMXPs) constitute a subclass of NS-LMXBs \citep{1998Natur.394..344W}. The spectra of AMXPs are often characterized by blackbody (BB) emission from a heated spot on the NS surface and Comptonization emission heated by the accretion shock \cite[e.g.,][]{2007A&A...464.1069F}.

As accreted hydrogen/helium matter accumulates on the NS surface, the accreted matter from the companion star is compressed and heated, occasionally leading to unstable burning if the temperature is sufficiently high, producing X-ray bursts. These bursts are known as type-\uppercase\expandafter{\romannumeral1} X-ray bursts \citep{2008ApJS..179..360G}.
Since the type-\uppercase\expandafter{\romannumeral1} X-ray bursts were discovered in the mid-1970s \citep{1976ApJ...206L.135B,1976ApJ...205L.127G}, thousands of such X-ray bursts have been observed in more than 100 NS-LMXBs \citep[e.g.,][]{2008ApJS..179..360G}. During the episodes of type-\uppercase\expandafter{\romannumeral1} X-ray bursts in the light curves of accreting NSs, the count rate increases abruptly in a short term ($\le$1-10 s), reaching a lot of times the persistent intensity, then decreases slowly in a relatively long term (from tens to hundreds of seconds). Generally, the type-\uppercase\expandafter{\romannumeral1} X-ray bursts show cooling tails, which are attributed to the NS photosphere cooling down after the fast injection of heat from nuclear reactions \citep[hereafter LCK11]{2011ApJ...733L..17L}. In some cases, quasi-periodic oscillations (QPOs) are found in X-ray bursts. Recently, \cite{2012ApJ...748...82L} argued that a mHz QPO is resulted from the thermonuclear burning when the mass accretion rate increases. \cite{2017A&A...606A.130I} found that the bolometric light curves during the decay phase in type-\uppercase\expandafter{\romannumeral1} X-ray bursts could be fit by the combination of a power law (PL) and a one-sided Gaussian function well. Type-\uppercase\expandafter{\romannumeral2} X-ray bursts were observed in MXB~1730-335 \citep{1993SSRv...62..223L} and GRO~J1744-28 \citep{1996Natur.379..799K}. It found that the inner disk radii are larger in the two sources than in other NS-LMXBs \citep{2014ApJ...796L...9D,2017MNRAS.466L..98V}. In some cases, the type-\uppercase\expandafter{\romannumeral2} X-ray bursts are defined as ones which lack cooling tails \citep[e.g.,][]{2002A&A...382..947K,2008ApJS..179..360G}. However, \citetalias{2011ApJ...733L..17L} found that the ratio of the peak-burst luminosity to the persistent luminosity could determine whether a cooling tail is present or not in a X-ray burst.

SAX~J1748.9-2021 was discovered by $\textit{BeppoSax}$ in 1998 \citep{1999A&A...345..100I}.
It is a NS X-ray transient located in the globular cluster NGC 6440 with a distance of $\sim$(8.2-8.5) kpc \citep{1994A&AS..108..653O,2003A&A...399..663K,2007AJ....133.1287V}. Until now, SAX~J1748.9-2021 has been observed in outburst fives times: 1998 \citep{1999A&A...345..100I}, 2001 \citep{2001ApJ...563L..41I}, 2005 \citep {2005ATel..495....1M}, 2010 \citep{2010ATel.2407....1P} and 2015 \citep{2015ATel.7106....1B}. The X-ray pulsations at around 442.3 Hz were detected during the 2001, 2005, 2010 and 2015 outbursts, from which an orbital period of $\sim$8.76 hours was inferred \citep{2007ApJ...669L..29G, 2008ApJ...674L..45A, 2016MNRAS.459.1340S}. Three type-\uppercase\expandafter{\romannumeral1} X-ray bursts were first detected during the 1998 outburst \citep{1999A&A...345..100I}, and then sixteen and four such X-ray bursts were observed in the 2001 \citep{2003ApJ...598..481K} and 2005 outbursts \citep{2008ApJ...674L..45A}, respectively. Recently, 25 type-\uppercase\expandafter{\romannumeral1} X-ray bursts were detected in the EPIC-pn data of {\it XMM-Newton} during the 2015 outburst \citep{2016MNRAS.457.2988P}. Assuming a pure-helium atmosphere, \cite{2008ApJS..179..360G} inferred a distance of 8.1 kpc to this source from the peak fluxes of its six Photospheric Radius Expansion (PRE) bursts. This AMXP hosts a nearly face-on accretion disk with a low inclination angle \cite[$\sim$8$^{\circ}$-14$^{\circ}$,][]{2017ApJ...844...53C}. A strong $\mathrm{H_{\alpha}}$ line was detected in this system, from which the mass, radius, and surface temperature of its companion star are estimated to be 0.70-0.83 $\mathrm{M_{\odot}}$, 0.88$\pm{0.02}$ $\mathrm{R_{\odot}}$, and 5250$\pm{80}$ K, respectively \citep{2017ApJ...844...53C}.  

In this work, we search and study the X-ray bursts of SAX~J1748.9-2021 during the 2010 outburst of this source, based on the observations on board \textit{Rossi X-ray Timing Explorer} (\textit{RXTE}). The observations and data analysis are described in Section 2, the results are presented in Section 3, and the discussions are given in Section 4. 

\section{Observations and data analysis}
\label{observation and analysis}

\begin{figure}
\includegraphics[width=1.0\linewidth,height=15cm]{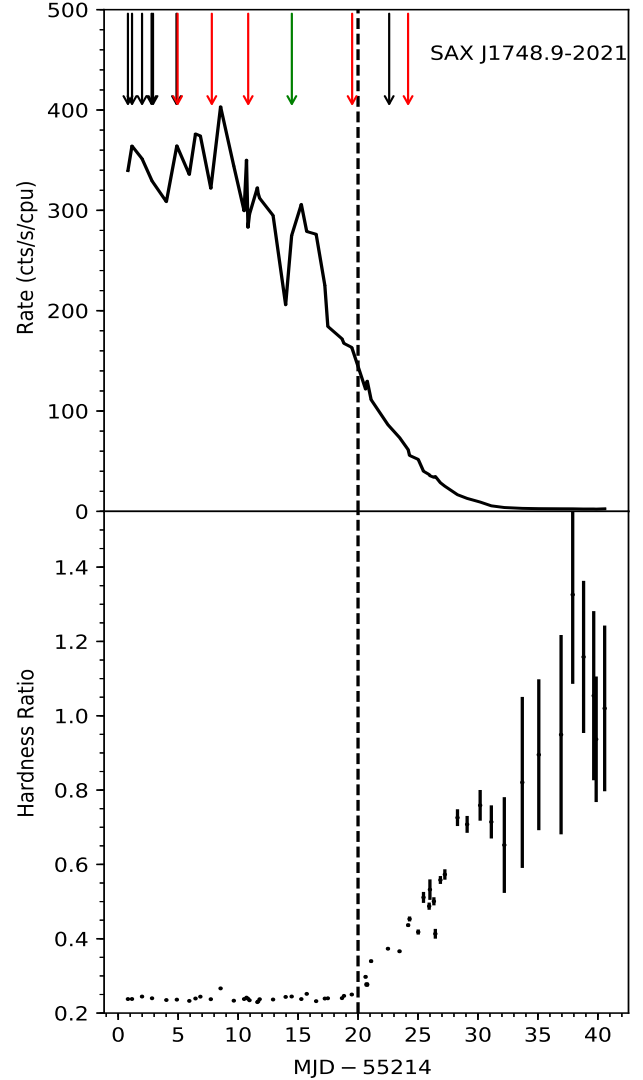}
\caption{The light curve of SAX~J1748.9-2021 during the 2010 outburst in 2-30 keV (top panel). The hardness ratio of 10-30 keV/5-10 keV is shown in the bottom panel. Each point in the diagram represents to the average count rate or hardness ratio of a single ObsID of $\emph{RXTE}$/PCA. The red and black spikes in the top panel mark the positions of the found 5 PRE X-ray bursts and 7 non-PRE X-ray bursts, respectively, and the green spike labels the location of the irregular X-ray burst. The vertical dashed line represents the time axis at MJD 55234.}
\label{fig:Flux}
\end{figure}

We use the publicly available data of SAX~J1748.9-2021 during its 2010 outburst on-board the $\emph{RXTE}$ satellite to perform our analysis. There are 59 observations of this outburst in the data set P94315 between Jan 18, 2010 and Feb 27, 2010 (MJD: from 55214 to 55254). The light curve of this outburst is shown in the top panel of Fig.~\ref{fig:Flux}. The exposure time of each observation spans a range of 0.43-10 ks and the exposure times during this outburst add up to $\sim$216 ks \citep{2016MNRAS.459.1340S}.  With HEASOFT 6.22, we perform our analysis using the data from the Proportional Counter Array (PCA), which was composed of five Proportional Counter Units (PCU0-4).

In this work, the Standard 2 mode data from PCU2, with time resolution of 16 s and energy band of 2-60 keV, are used to extract the spectra of persistent emission by use of the FTOOLS SAEXTRCT. Usually, a spectrum of persistent emission is produced with the data of 160 s prior to a X-ray burst. Manipulating FTOOLS PCABACKEST, we produce the PCA background files from the PCA background model provided by {\it RXTE} team and, then, extract the PCA background spectra with SAEXTRCT. When extracting spectra, we chose the good data by inputting the Good Time Interval (GTI) files, which are created operating FTOOLS MAKETIME with the criteria of elevation angle > 10$^\circ$ and pointing offset < 0.02$^\circ$. The response matrices are produced with PCARSP V11.7.1, a perl script of {\it RXTE}. With the response matrices, we construct a combination model consisting of a BB (BB1), a PL, plus a line component taking the form of GAUSS to fit the background-subtracted PCA spectra of persistent emission. In the combination model of BB1+PL+GAUSS, the BB1 describes the NS emission due to accretion, the PL interprets the emission from the accretion disk, and the GAUSS accounts for the iron lines. It is noted that this model is similar to the so-called ``Western Model   '' \citep{White1986}, which has been frequently used to fit the spectra of NS-LMXBs. 

The Event mode data (E\_125us\_64M\_0\_1s) of PCU2 are used to search and analyze the X-ray bursts within this outburst. Firstly, in order to search X-ray bursts, we produce light curves with time bin of 1 s and in 2-60 keV with FTOOLS SEEXTRCT. We find 13 X-ray bursts, which are listed in Table~\ref{tab:table info} and shown in Fig.~\ref{fig:lightcurve-paper}. The start time of a X-ray burst, listed in Table~\ref{tab:table info}, is defined as the time point when the count rate rises to 10\% of the peak count rate in the rise phase of the X-ray burst. Fig.~\ref{fig:hardness-paper} shows the hardness ratio of the detected X-ray bursts, which is defined as the ratio of the count rate in 6-30 keV to the count rate in 2-6 keV. In Figs.~\ref{fig:lightcurve-paper} and \ref{fig:hardness-paper}, the coordinate origin in the horizontal axis is defined as the time point when the intensity reaches the peak count rate. Secondly, with time bin of 0.25 s, we extract the time-resolved spectra of every X-ray burst by use of SEEXTRCT. Finally, we construct a model of BB2+$f_{a}$(BB1+PL+GAUSS) to fit the spectra of all the X-ray bursts. In this model, the BB2 accounts for the emission of the X-ray bursts due to the thermonuclear burning on the NS surface. The combination model of BB1+PL+GAUSS describes the persistent emission during the X-ray bursts. Taking into account the variation of the persistent emission during these X-ray bursts, this combination model is modified with a multiplicative factor $f_{\rm a}$ \citep{2013ApJ...772...94W,2015ApJ...801...60W}.

The spectra of persistent emission and the spectra during the X-ray bursts are fit with $\textsc{xspec}$ 12.9.1 m \citep{1996ASPC..101...17A} in 3-20 keV. When fitting, a multiplicative model (PHABS in XSPEC) describing interstellar absorption \citep{2000ApJ...542..914W} is used, a systematic error of 0.5\% is added into the spectra due to calibration uncertainties, the parameters of the model interpreting the persistent emission during the X-ray bursts are fixed at those during the non X-ray bursts, and the parameters are estimated within the confidence level of 90\%. The line energy and line width of the iron line are fixed at 6.7 keV and 0.4 keV, respectively \citep{2016MNRAS.457.2988P}. Because the BB temperature is less than several keV, the BB flux in the energy band less than 3 keV contributes much to the total flux of BB, so we apply the convolution model of CFLUX to estimate the flux in 1.5-30 keV. 

After fitting all the spectra of each X-ray burst, we get its time-resolved flux spectrum. Following the method of \cite{2017A&A...606A.130I}, we obtain the delay time of each X-ray burst by fitting the descending segment of its time-resolved flux spectrum with a two-component model consisting of a PL and a one-sided Gaussian function:
\begin{equation}
\mathrm{F(t) = F(t_0) \left( \frac{t}{t_0} \right)^{-\alpha} + \frac{G}{\sqrt{2\pi}s} \; {\rm e}^{-\frac{(t-t_{s})^2}{2s^2}}}
\end{equation}
where G is the normalization of the Gaussian function and s is the standard deviation of flux spectrum. The Gaussian function centroid is fixed to the start of the burst ($\mathrm{t_{s}}$) and $\mathrm{t_{0}}$ is defined as the time point when the flux descends below 55\% of the peak flux. The end time of the X-ray burst is defined as the time point when the flux drops down to 10\% of the peak flux. It is noted that the error bars are not considered in this analysis, because the flux values fluctuate remarkablly. Then, we measure the fluence of each X-ray burst ($\mathrm{E_{b}}$) in 1.5-30 keV by integrating over the complete duration of this X-ray burst, which is listed in Table~\ref{tab:table info}.

\section{Results}

\subsection{Outburst}

Excluding the count rates during the X-ray bursts, the hardness ratio of the outburst is defined as the count rate ratio of 10-30 keV/5-10 keV, which is shown in the bottom panel of Fig.~\ref{fig:Flux}. 
The decay phase of the 2010 outburst from SAX J1748.9-2021 lasted about 30 days, which is similar to the 2001, 2005 and 2015 outbursts \citep{2008ApJ...674L..45A,2016MNRAS.457.2988P}.  
As shown in Fig.~\ref{fig:Flux}, after MJD 55234, the count rate in 2-30 keV decreased remarkably, while the hardness ratio increased notably, which might suggest that this source undergoes a spectral transition.

\begin{table*}
\small
\newcommand{\tabincell}[2]{\begin{tabular}{@{}#1@{}}#2\end{tabular}}
\centering
\caption{The properties of the 13 X-ray bursts during the 2010 outburst in SAX~J1748.9-2021. }
\label{tab:table info}
\begin{tabular}{lcccccccccccc}
        \hline
		No & OBSID$^{a}$ & \tabincell{c}{Start $\mathrm{T_{start}^{b}}$\\(UT,2010)} & \tabincell{c}{Intensity$^{c}$\\($\mathrm{cts\,s^{-1}}$)} & \tabincell{c}{$\mathrm{F_{per}}$$^{d}$ \\ ($10^{-9})$}  & \tabincell{c}{$\mathrm{F_{l}}$$^{e}$ \\ ($10^{-9})$}  &$f_{\rm a}$$^{e}$  & \tabincell{c} {$\mathrm{E_{b}}$$^{f}$ \\$(10^{-8})$}  & $\beta$$^{g}$\\
		\hline
B1  & *-05-00 & Jan 19 04:20:42 &2009 &4.34$_{-0.12}^{+0.15}$ &22.39$_{-3.64}^{+3.92}$& 0.87$\pm{0.53}$ &21.8$_{-1.2}^{+1.2}$ & $5.2_{-1.0}^{+1.1}$ \\
B2  & *-05-01 & Jan 18 19:23:10 &1741 &4.12$_{-0.11}^{+0.14}$ &20.68$_{-3.82}^{+2.53}$& 0.35$\pm{0.58}$ &21.8$_{-1.3}^{+1.3}$ & $5.0_{-1.1}^{+0.8}$  \\
B3  & *-05-02 & Jan 19 23:46:23 &2279 &4.25$_{-0.13}^{+0.16}$ &21.57$_{-3.24}^{+3.55}$& 1.94$\pm{0.51}$ &19.5$_{-0.7}^{+0.8}$ & $5.1_{-0.9}^{+1.0}$ \\
B4  & *-05-03 & Jan 20 18:51:15 &2066 &4.22$_{-0.11}^{+0.09}$ &21.27$_{-3.48}^{+3.69}$& 0.82$\pm{0.56}$ &22.5$_{-0.8}^{+0.9}$ & $5.0_{-1.0}^{+1.0}$ \\
B5  & *-05-03 & Jan 20 21:44:22 &1850 &4.06$_{-0.10}^{+0.12}$ &18.41$_{-4.10}^{+4.24}$& 0.93$\pm{0.67}$ &26.0$_{-0.9}^{+0.9}$ & $4.5_{-1.1}^{+1.2}$ \\
B6  & *-06-01 & Jan 22 20:40:24 &1900 &4.29$_{-0.08}^{+0.09}$ &25.17$_{-4.18}^{+2.76}$& 0.34$\pm{0.56}$ &22.6$_{-0.8}^{+0.8}$ & $5.9_{-1.1}^{+0.8}$ \\
B7  & *-06-01 & Jan 22 22:53:56 &2635 &4.33$_{-0.09}^{+0.11}$ &26.82$_{-4.66}^{+5.02}$& 1.55$\pm{0.66}$ &18.3$_{-1.0}^{+1.0}$ & $6.2_{-1.2}^{+1.3}$  \\
B8  & *-06-05 & Jan 25 19:24:06 &3425 &3.79$_{-0.08}^{+0.09}$ &43.41$_{-3.59}^{+3.71}$& 1.73$\pm{0.56}$ &13.0$_{-0.9}^{+0.9}$  & $11.4_{-1.2}^{+1.3}$ \\
B9 & *-06-11 & Jan 28 20:15:51 &3270 &4.08$_{-0.08}^{+0.09}$ &35.22$_{-4.86}^{+5.34}$& 2.51$\pm{1.43}$  &14.9$_{-0.9}^{+1.0}$ & $8.6_{-1.4}^{+1.5}$ \\
B10 & *-07-02 & Feb 01 11:31:51 &665  &3.31$_{-0.09}^{+0.11}$ & 5.29$_{-2.37}^{+2.44}$& 0.75$\pm{0.66}$ &1.0$_{-0.2}^{+0.2}$ & $1.6_{-0.8}^{+0.8}$  \\
B11 & *-08-01 & Feb 06 12:12:26 &3355 &2.14$_{-0.10}^{+0.13}$ &43.10$_{-3.34}^{+3.42}$& 1.87$\pm{1.11}$ &24.1$_{-0.8}^{+0.8}$ & $20.1_{-2.8}^{+2.5}$ \\
B12 & *-08-05 & Feb 09 14:19:45 &1765 &1.24$_{-0.05}^{+0.07}$ &25.94$_{-3.45}^{+3.60}$&-3.90$\pm{2.19}$ &13.9$_{-0.9}^{+0.9}$ & $20.9_{-3.6}^{+4.1}$ \\
B13 & *-08-07 & Feb 11 04:09:24 &2636 &0.91$_{-0.04}^{+0.05}$ &40.93$_{-3.97}^{+4.11}$&-3.95$\pm{3.57}$ & 20.4$_{-0.7}^{+0.7}$ & $45.0_{-6.3}^{+7.0}$ \\
        \hline
\end{tabular}

\begin{enumerate}
 \item[$^a$]{'${\mathrm{*}}$' stand for 94315-01.}
 \item[$^b$]{The start time is defined as the first point is larger than 10\% of the largest count rate in light curve in the time bin of 1 s.}
 \item[$^c$]{Bursts peak count rate in the 2.0-60.0 keV energy band of PCU 2. The background is subtracted.}
 \item[$^d$]{Persistent emission flux in the unit of $\mathrm{erg~cm^{-2}~s^{-1}}$ at the 1.5-30.0 keV energy band of PCU 2. The background is subtracted.}
 \item[$^e$]{The largest bolometric flux and corresponding $f_{\rm a}$ in the time-resolved spectroscopy of each X-ray burst.}
 \item[$^f$]{Burst fluence in the unit of $\mathrm{erg~cm^{-2}}$ at 1.5-30.0 keV.}
 \item[$^g$]{$\mathrm{\beta\,\equiv\,F_{l}/F_{per}}$.}
\end{enumerate}
\end{table*}

\begin{figure*}
\centering
\includegraphics[width=1\linewidth]{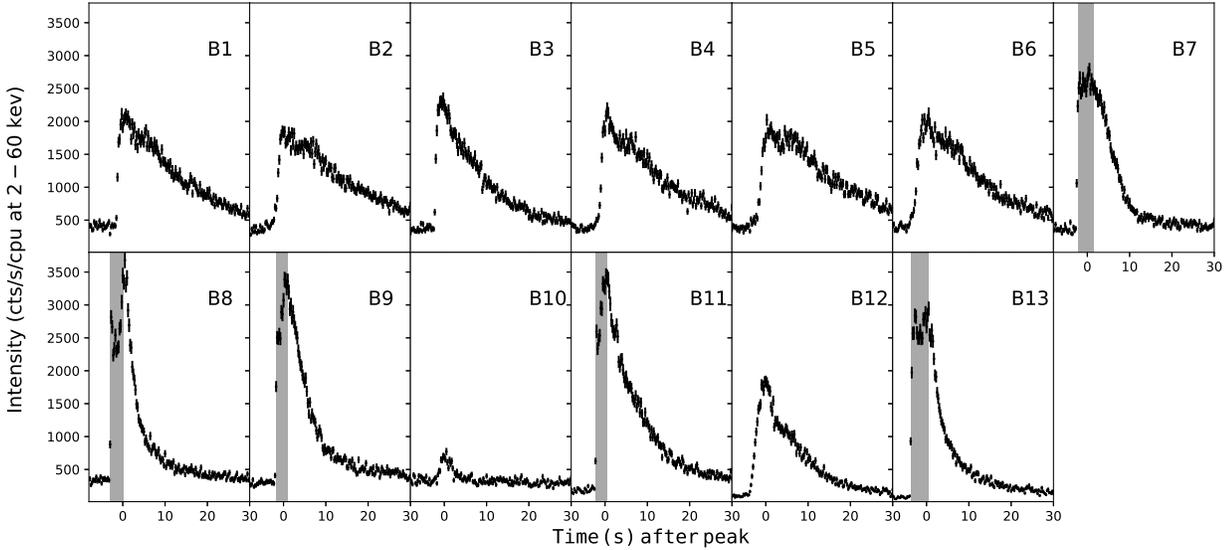}\hfill
\caption{The background-subtracted light curve at 2.0-60.0 keV of the 13 bursts. The time resolution is 0.25 s. The gray areas represent the PRE phases of the 5 PRE X-ray bursts. Time zero on the X-axis is the time point when the intensity reaches the peak count rate (see Table~\ref{tab:table info}).}
\label{fig:lightcurve-paper}
\end{figure*}

\begin{figure*}
\centering
\includegraphics[width=1\linewidth]{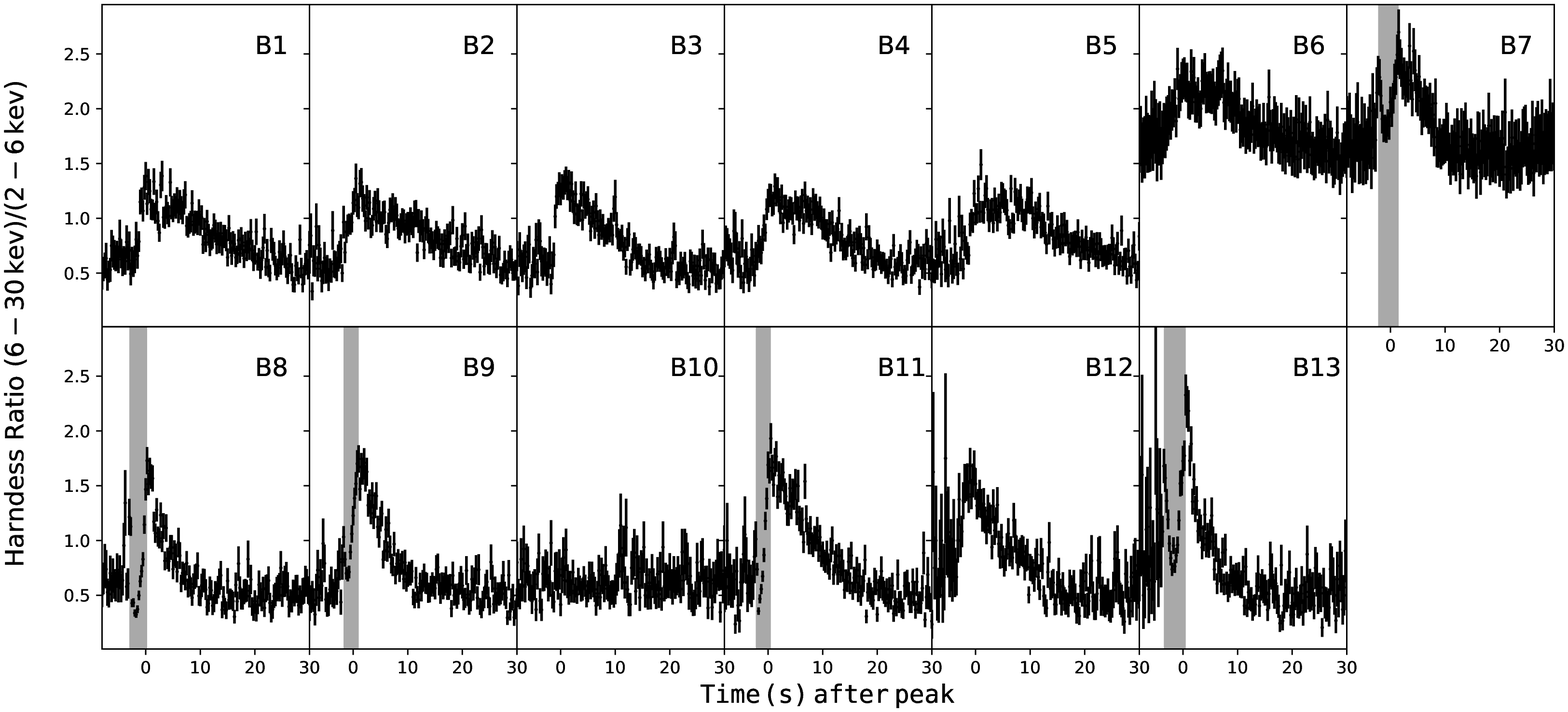}\hfill
\caption{Hardness ratios (6-30 keV)/(2-6 keV) of the 13 bursts. The time resolution is 0.25 s. The gray areas represent the PRE phase. Time zero on the X-axis is the time point when the intensity reaches the peak count rate (see Table~\ref{tab:table info}).}
\label{fig:hardness-paper}
\end{figure*}

\subsection{X-ray bursts}

In this work, we find 13 X-ray bursts during the 2010 outburst of SAX~J1748.9-2021, which are listed in Table~\ref{tab:table info} and shown in Fig.~\ref{fig:lightcurve-paper}. Moreover, these detected X-ray bursts are marked with spikes during the outburst in the top panel of Fig.~\ref{fig:Flux}, showing that X-ray bursts \#1-11 occur when the hardness ratios are $\sim$0.25, nevertheless X-ray bursts \#12-13 are present when the hardness ratios are $\sim$0.37 and $\sim$0.44, respectively. It is noted that the count rate is much smaller than of other X-ray bursts and all the other X-ray bursts show cooling tails, but the X-ray burst \#10 lacks such a cooling tail. Moreover, the duration of the X-ray bursts except the X-ray burst \#10 spans a range of 10-30 s. Therefore, all the detected X-ray bursts are the type-\uppercase\expandafter{\romannumeral1} X-ray bursts except the  X-ray burst \#10 (\citetalias{2011ApJ...733L..17L}). 

The hardness ratio can be used to investigate the X-ray burst evolution \citep{2006csxs.book..113S,2010A&A...510A..81C}. As shown in Fig.~\ref{fig:hardness-paper}, except burst \#10, the hardness ratio of all the X-ray bursts shows remarkable evolution. It is noted that within the rise phase of bursts \#7, 8, 9, 11 and 13, the hardness ratio of these bursts decreases abruptly and, then, increases rapidly, which demonstrates the phase for PRE phenomenon. All the observed PRE X-ray bursts reach their PRE phase within 0.5 s, as shown in Figs.~\ref{fig:lightcurve-paper} and \ref{fig:hardness-paper}. The red spikes in the top panel of Fig.~\ref{fig:Flux} mark the 5 PRE X-ray bursts in the outburst, and the black spikes in this panel mark the 7 non-PRE X-ray bursts, i.e. the X-ray bursts \#1-6 and 12. 

\begin{figure*}
\centering
     \includegraphics[width=0.45\linewidth,height=6cm]{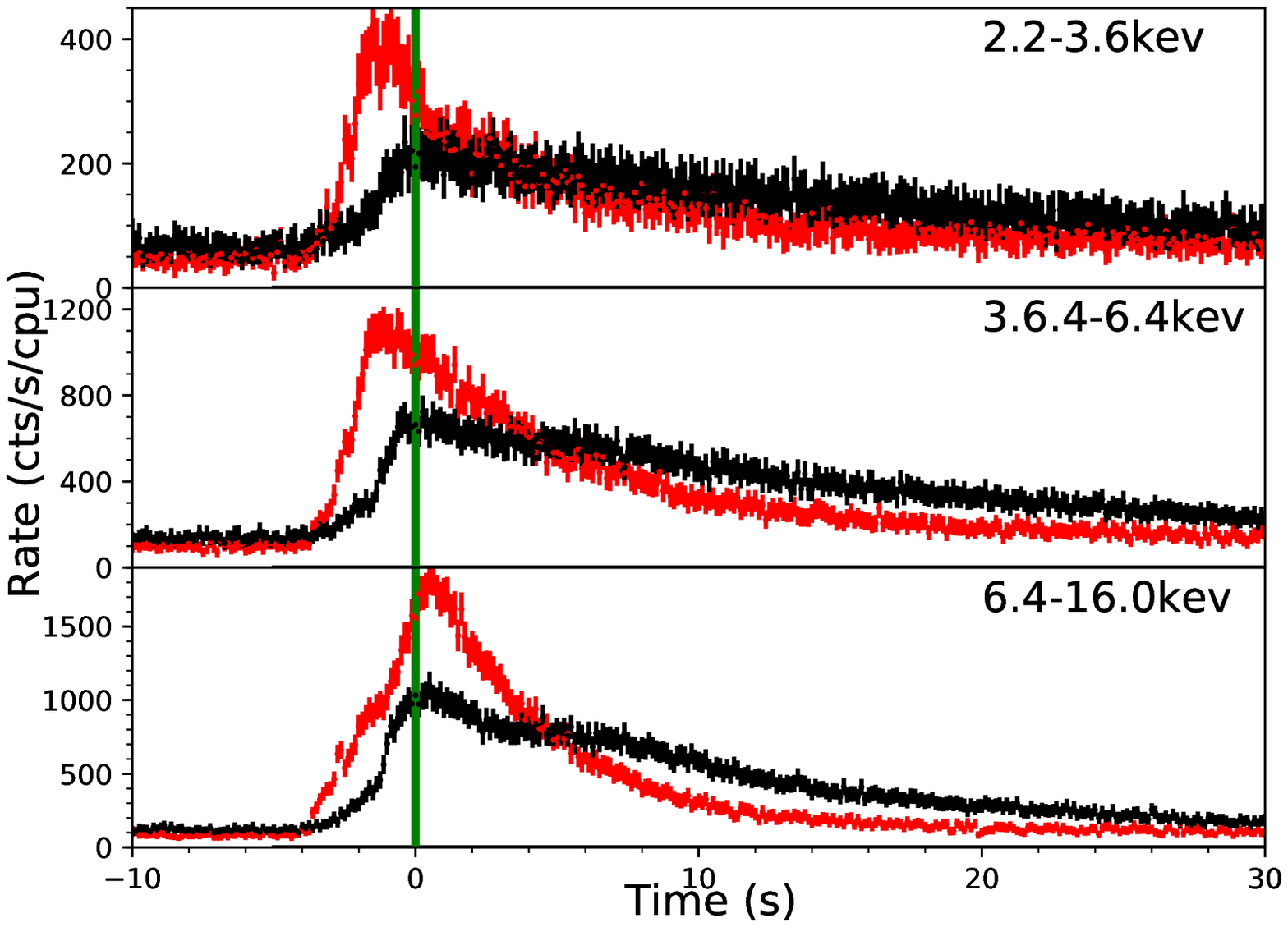}\hfill
     \includegraphics[width=0.45\linewidth,height=6cm]{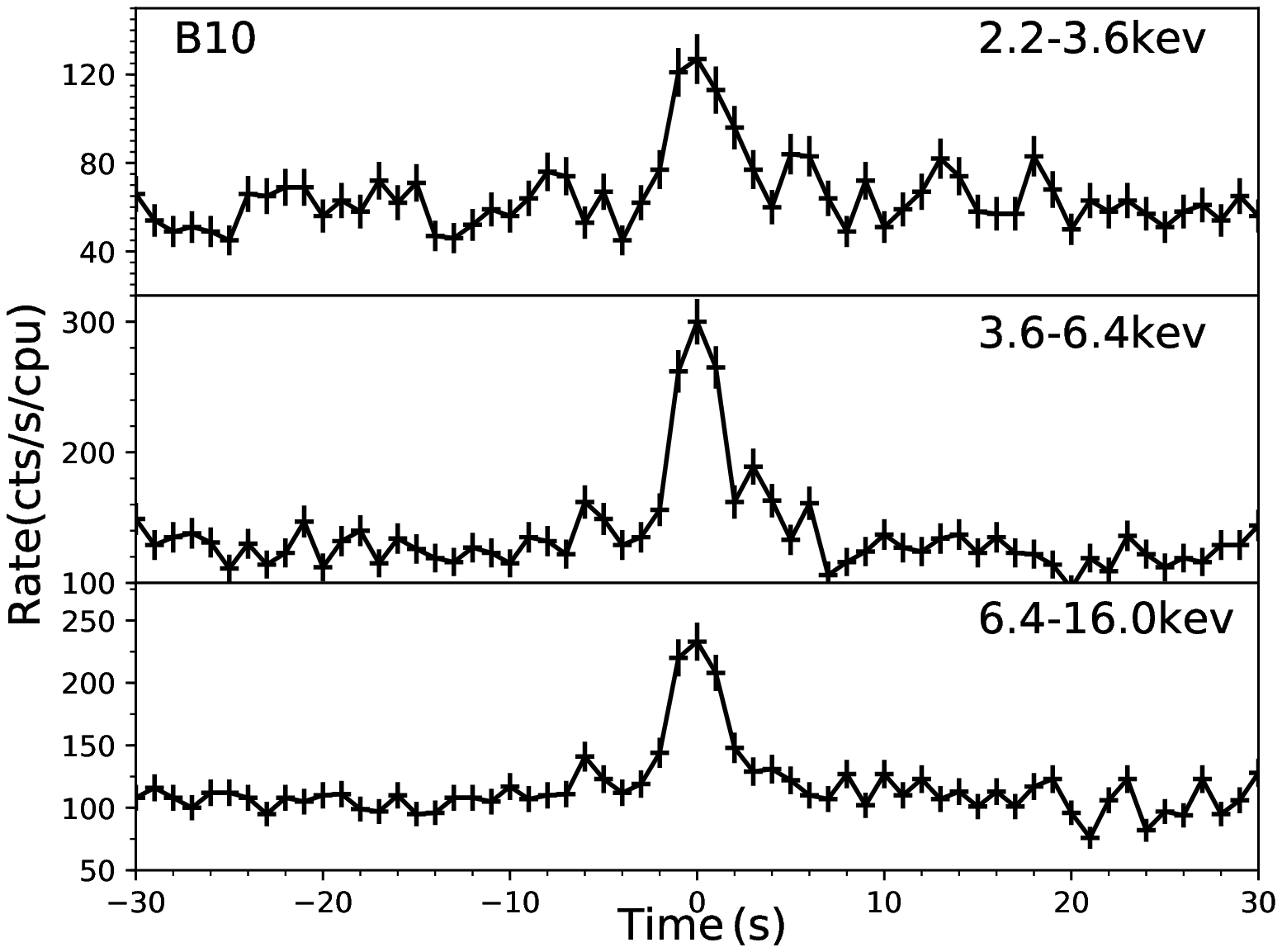}\hfill
     \caption{\emph{Left panel} - Average light curves for non-PRE and PRE bursts in three energy bands with a time resolution of 0.125 s. The black and red curves represent the non-PRE and PRE bursts, respectively. The green line represents the time axis at 0 s. The light curves are corrected for background emission.
     \emph{Right panel} - the light curves of burst \#10 in three energy bands, with time bin of 1 s. The background emission is subtracted.}
     \label{fig:average-lightcurve}
\end{figure*}

As shown in Fig.~\ref{fig:hardness-paper}, the duration is obviously larger of the non-PRE X-ray bursts than of the PRE X-ray bursts. It is noted that the duration of the X-ray burst \#12 is the shortest among the non-PRE X-ray bursts. In order to further compare the two types of the found type-\uppercase\expandafter{\romannumeral1} X-ray bursts, we calculate the average light curves of the PRE X-ray bursts and non-PRE X-ray bursts in three energy bands, i.e. 2.1-3.6 keV, 3.6-6.4 keV, 6.4-16.0 keV, respectively, which are displayed in the left panel of Fig.~\ref{fig:average-lightcurve}. It is shown that in any energy band the average peak count rate is larger of the PRE bursts than of the non-PRE bursts, while the burst profile is wider of the non-PRE bursts than of the PRE bursts. The average peak count rate of the PRE bursts or non-PRE bursts increases when the energy band goes up. In any energy band, the peak count rate of the non-PRE bursts occurs at the coordinate origin of the horizontal axis, while the peak count rate of the PRE bursts moves left when the energy band goes down.  

We would like to make a comparison between the X-ray burst \#10 and the standard type-\uppercase\expandafter{\romannumeral1} X-ray bursts. The light curves of this burst in the three energy bands are shown in the right panel of Fig.~\ref{fig:average-lightcurve}. It is seen that among the three energy bands the largest peak count rate of this burst occurs in the middle energy band, while the largest peak count rate of the average light curve of the PRE or non-PRE bursts occurs in the highest energy band. The peak point of this burst does not move horizontally, which is similar to behavior of the non-PRE bursts, but the peak point of the PRE burst moves left, as pointed above. The profile of the X-ray burst \#10 is obviously narrower than that of the PRE or non-PRE bursts. Moreover, as mentioned above, this burst lacks a cooling tail, but the PRE or non-PRE bursts show obvious cooling tails. Therefore, in this paper we call the X-ray burst \#10 an irregular X-ray burst, which is marked with the green spike in the top panel of Fig.~\ref{fig:Flux}.   

\begin{figure}
\centering
\flushleft
     \includegraphics[width=0.9\linewidth]{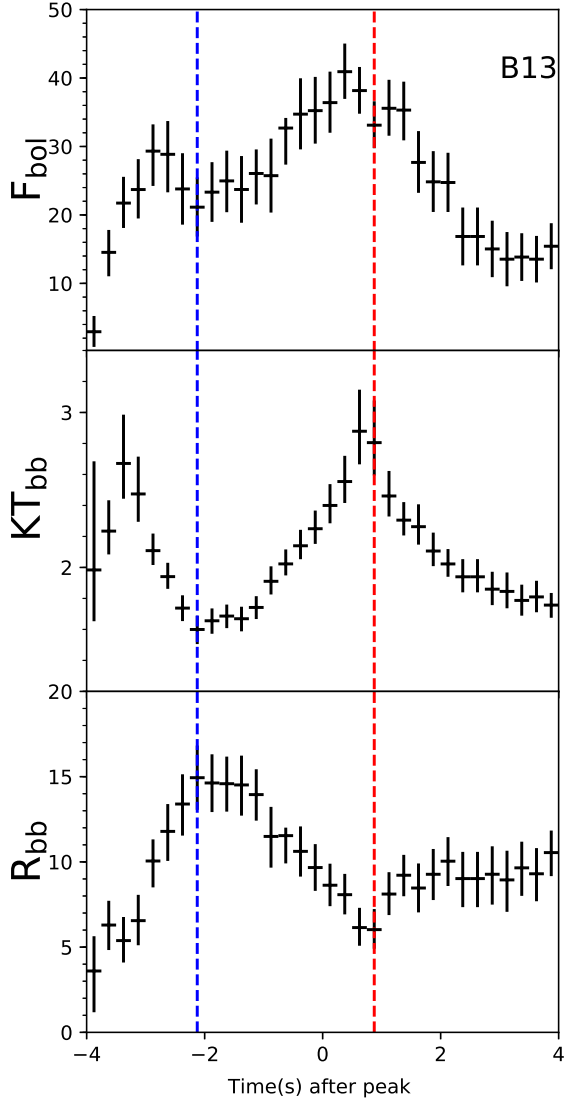}\\
     \caption{Spectral evolution of burst \#13. From top to bottom: bolometric flux estimated in 1.5-30 keV, temperature, radius of the photosphere obtained by fitting the burst spectra with a \textsc{BB} model ($\textsc{xspec}$). The red and blue dashed lines show the ``touch-down'' and PRE phenomena, respectively.}
     \label{fig:touch down}
\end{figure}

\begin{figure}
\flushleft
     \includegraphics[width=1\linewidth]{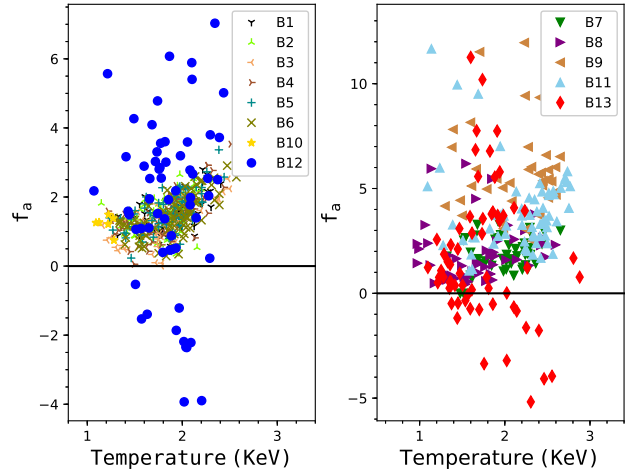}\\
     \caption{The correlation between the multiplicative factor $f_{a}$ and the X-ray burst temperature. The left and right panels show the results of the non-PRE X-ray bursts as well as the irregular X-ray burst and the PRE X-ray bursts, respectively.}
     \label{fig:A-T}
\end{figure}

\begin{figure}
\centering
\flushleft
     \includegraphics[width=0.9\linewidth]{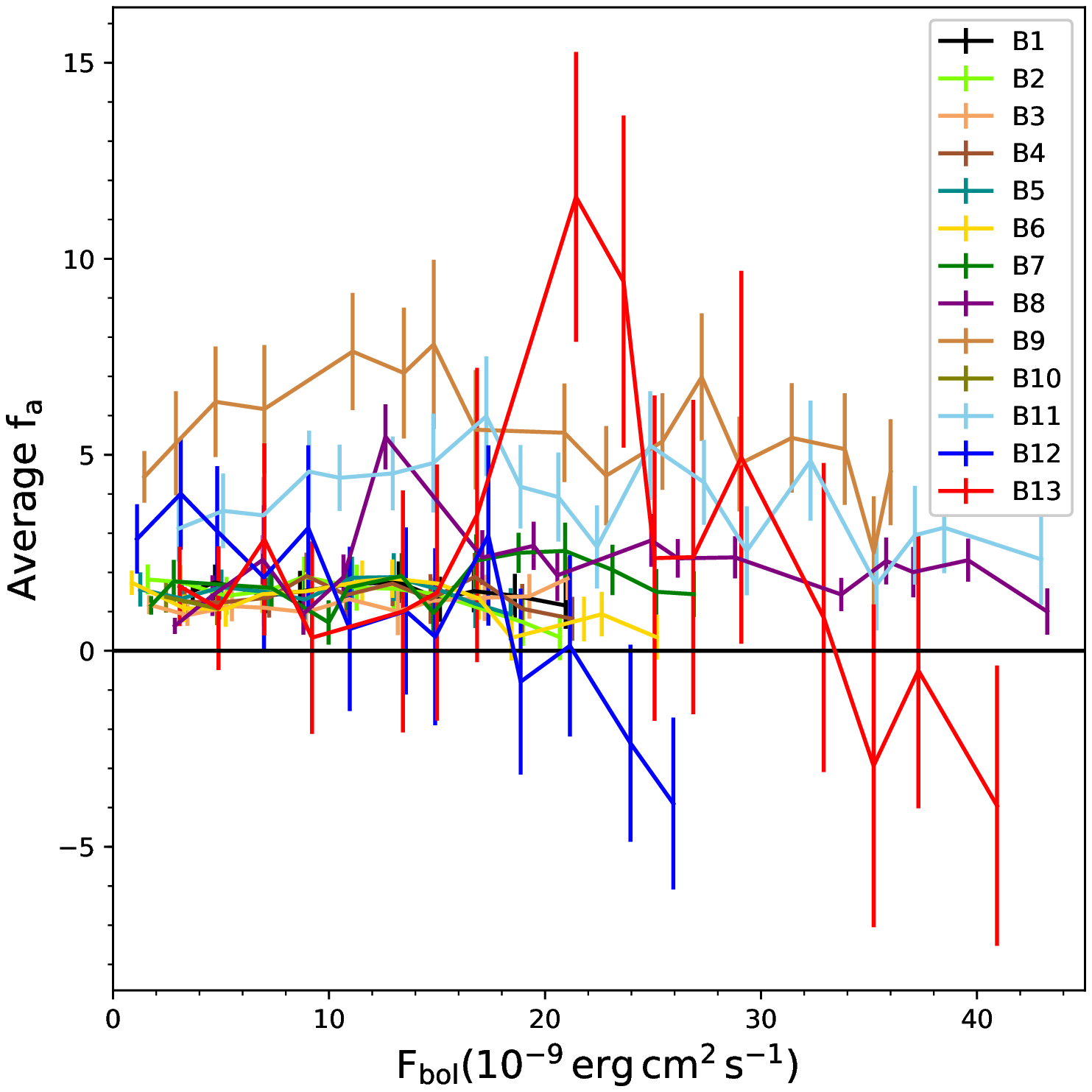}\\
     \caption{Flux vs. the multiplicative factor $f_{a}$. The horizontal black line shows when the average $f_{a}$ = 0.}
     \label{fig:A-F}
\end{figure}

\subsection{Spectral fitting}

In this work, a combination model of BB1+PL+GAUSS is applied to fit the spectra of persistent emission, while the model of BB2+$f_{a}$(BB1+PL+GAUSS), another combination model, is used to fit the spectra during the X-ray bursts. The value of $\chi^{2}_{\nu}$ varies in the range of $\sim$0.8-1.2, indicating that the fittings are acceptable statistically well. From the spectral fitting, we obtain  the flux of persistent emission ($\rm {F_{per}}$), the peak flux of every burst ($\rm {F_{l}}$), and the parameter $\beta$ ($\beta$ $\equiv$ $\mathrm{F_{l}/F_{per}}$), which are listed in Table~\ref{tab:table info}. We obtain the fluence of every X-ray burst ($\rm {E_{b}}$) through fitting its time-resolved flux spectrum with Equation 1. The obtained value of $\rm {E_{b}}$ and the value of the multiplicative factor $f_{\rm a}$ of the peak flux of every X-ray burst are also listed in Table~\ref{tab:table info}. Inspecting the evolution of the source intensity with the detected X-ray bursts in Fig.~\ref{fig:Flux} and the values of $\beta$ listed in the last column of Table~\ref{tab:table info}, it is concluded that generally, the value of $\beta$ increases with decreasing of $\dot{\rm M}$, but the value of $\beta$ of the irregular X-ray burst is the least among the 13 X-ray bursts, although it occurs in episode of intermediate $\dot{\rm M}$ during the outburst. 

From spectral fitting, we obtain the evolution of the parameters of BB2, i.e. the parameters describing the emission due to the thermonuclear burning on the NS surface. Fig.~\ref{fig:touch down} shows the evolution of the BB2 parameters of the X-ray burst \#13. In this figure, it is shown that around the blue line, the BB temperature reaches the minimum, while the BB radius ($\rm R_{bb}$) reaches the maximum, being larger than the NS radius (10 km) and, therefore, showing the typical PRE phenomenon. The X-ray burst \#13 is a peculiar X-ray burst. Not only is it a PRE X-ray burst, but also it displays the so-called ``touch-down'' behavior, which is shown around the red line in Fig.~\ref{fig:touch down}: the BB temperature ($\rm kT_{bb}$) reaches the maximum, while the BB radius ($\rm R_{bb}$) touches the minimum, and the burst flux ($\rm F_{bol}$) approaches the maximum.

\subsection{The behavior of the multiplicative factor}

The evolution of the multiplicative factor ($f_{\rm a}$) with the BB temperature ($\rm {kT_{BB}}$) of the non-PRE as well as the irregular X-ray burst and the PRE X-ray bursts, is shown in the left and right panels of Fig.~\ref{fig:A-T}, respectively. Except the non-PRE X-ray burst \#12 and the PRE X-ray burst \#13, it is shown that $f_{\rm a}$ is positively correlated with $\rm {kT_{BB}}$ for the non-PRE X-ray bursts and the points of these non-PRE X-ray bursts are relatively concentrated, but obvious correlation between $f_{\rm a}$ and $\rm {kT_{BB}}$ cannot be seen for the PRE X-ray bursts and the points of these PRE X-ray bursts are distributed somewhat randomly. It is noted that any correlation between the two parameters cannot be seen for the burst \#12 or \#13 and the points of any of the two bursts are distributed very randomly and diffusely. Moreover, all the values of $f_{\rm a}$ of all the X-ray bursts are positive except the bursts \#12 and \#13, while they are negative of the bursts \#12 and \#13 in some cases. It is important to point out that the two X-ray bursts, i.e. \#12 and \#13 are detected in the episodes with relatively low $\dot{\rm M}$ during the outburst, as sown in Fig.~\ref{fig:Flux}, might implying that the two bursts are in the hard state of this source. The negative values of $f_{\rm a}$ were also found in the hard state of atoll source 4U 1608-52 \citep{2014ApJ...791L..39J}. 

In order to further investigate the behavior of the multiplicative factor, we calculate the average value of $f_{\rm a}$ in each flux bin of 2$\times10^{-9}$ $\mathrm{erg\,cm^{-2}\,s^{-1}}$ and the evolution of $f_{\rm a}$ with the burst flux is shown in Fig.~\ref{fig:A-F}. Any correlation between the multiplicative factor and the burst flux cannot be seen for the detected X-ray bursts except the  bursts \#12 and \#13. However, for the X-ray burst \#12 or \#13, when the burst flux is larger than a critical value, i.e. $\sim$2.1$\times10^{-8}$ $\mathrm{erg\,cm^{-2}\,s^{-1}}$, the anti-correlation between the two parameters is visible and the negative values of $f_{\rm a}$ are present in some cases.

\section{Discussion}

In this work, we find 12 standard type-\uppercase\expandafter{\romannumeral1} X-ray bursts, including 7 non-PRE X-ray bursts and 5 PRE X-ray bursts, and an irregular X-ray burst. As shown by Fig.~\ref{fig:Flux}, Most of the X-ray bursts (\#1-9) are found in the episode with relatively high $\dot{\rm M}$ during the outburst, two X-ray bursts (\#10-11) are detected in the term with the intermediate $\dot{\rm M}$, and the last two X-ray bursts (\#12-13) are present in the duration with relatively low $\dot{\rm M}$. The X-ray bursts disappear when the $\dot{\rm M}$ of the source is extremely low. Most of the non-PRE X-ray bursts are found in the episode with relatively high $\dot{\rm M}$, while the occurrence of the PRE X-ray bursts does not depend on $\dot{\rm M}$.

\subsection{Touch down}

For the PRE X-ray bursts, the radius of the photosphere could be much larger than the NS radius during the expansion episode, while the photosphere will contract during the contraction phase and its radius might reach the minimum, i.e the NS radius, which is called the so-called ``touch-down'' phenomenon \citep{Damen1990}. In this work, we find such a PRE X-ray burst, i.e. the PRE X-ray burst \# 13, as shown in Fig.~\ref{fig:touch down}. \cite{2014MNRAS.445.4218K} suggested that only the PRE X-ray bursts in the hard state could show such ``touch-down'' behavior. As pointed out above, the X-ray burst \#13 is found in the episode with relatively large hardness ratio and meanwhile, relatively low $\dot{\rm M}$, might suggesting that this X-ray burst is in the hard state of this source. Moreover, the value of the multiplicative factor ($f_{\rm a}$) of this PRE X-ray burst is negative occasionally, suggesting that the accreting matter may be thrown out \citep{2014ApJ...791L..39J}, which could happen when the inner disk radius is relatively large, being the typical characteristic of the hard state.    

The ``touch-down'' phenomenon can be a way through which to determine the NS mass and radius \citep{Damen1990,Ozel2009}. Here, with the spectral parameters of the touchdown, i.e. $\rm {kT_{bb}}$ ($\sim$ 2.8$_{-0.2}^{+0.3}$ keV), burst flux (33.1$_{-3.4}^{+3.5}$$\times$ 10$^{-9}$ erg cm$^{-2}$ s$^{-1}$), and $\mathrm{R_{bb}}$ ($\sim$ 6.0$_{-1.1}^{+1.2}$ km), we estimate the NS radius as follow \citep{1993SSRv...62..223L} 
\begin{equation}
 \mathrm{R_{NS} = \frac{R_{\rm bb}}{(1+z)}*f_{\rm c}^{2}(td)}
\end{equation}
where $\rm {f_{c}(td)}$ is the colour correction factor at the time of touchdown (td). Since the inner disk truncation radius is large enough in the hard state, so the entire NS surface may be visible. Therefore, the anisotropy of X-ray emission will not be taken account of for us to derive the NS radius ($\mathrm{R_{NS}}$). The burst flux approaches the maximum at td and then drops to the half of the touch down flux at td/2. All NS atmosphere models suggest that the color-correction factor ($\rm {f_{c}}$) decreases in the beginning of cooling phase when the X-ray burst luminosity drops below the Eddington value \citep{2011A&A...527A.139S}. Therefore, we assume $\rm {f_{c}}$ to be a relatively low value at td/2, i.e. $\rm {f_{c}(td/2)}\approx$1.4, and derive the value of $\rm {f_{c}(td)}$\citep{2014MNRAS.445.4218K}. With the inferred $\rm {f_{c}(td)}$, the value of $\mathrm{R_{bb}}$, and taking the value of gravitational redshift (z) to be 0.31, we obtain $\rm {R_{NS}}\approx$13.2$_{-2.4}^{+2.6}$ km. The derived value of $\mathrm{R_{NS}}$ approximates to the values of 8.18$\pm$1.62 km and 10.93$\pm$2.09 km reported in the paper of \cite{2013ApJ...765L...1G}. However, the inferred value of $\mathrm{R_{NS}}$ is somehow larger than that reported by \cite{2016MNRAS.457.2988P}, i.e. 7.0-7.6 km, which might remind us to consider the anisotropic X-ray emission for deriving $\mathrm{R_{NS}}$.

It is noted that the X-ray burst flux at the touchdown just approaches the peak flux, so the touchdown flux and the peak flux can be compared each other. \cite{2008MNRAS.387..268G} suggested that if the ratio of the peak flux to the touchdown flux ($\mathrm{f=f_{l}/f_{td}}$) is smaller than 1.6, the source will possess a low inclination angle and then, it will be a non-dip source. For SAX J1748.9-2021, this ratio is $\sim$1.24$_{-0.25}^{+0.26}$, indicating that this source hosts a nearly face-on disk. 

\subsection{Persistent emission}

The persistent emission due to accretion could be critical for X-ray bursts to occur and it might be responsible for the characteristics of X-ray bursts. \cite{2007ApJ...654.1022P} suggested that the unstable burning of hydrogen can trigger the burning of a thick layer of helium when the mass accretion rate is below 0.3\%$\rm {\dot{M}_{edd} }$. Until now, some type-\uppercase\expandafter{\romannumeral1} X-ray bursts with relatively long duration (tens of minutes) have been observed at such low mass accretion rate of persistent emission \citep{2005A&A...441..675I,2008A&A...484...43F,2015ApJ...799..123B,2017ApJ...836..111K}. However, \cite{2006ApJ...648L.123C} proposed that the type-\uppercase\expandafter{\romannumeral1} X-ray bursts will occur only when the accretion rate is below a critical value, i.e. $\sim$0.3$\mathrm{\dot{M}_{\rm edd}}$, which is conformed with the $\gamma-\tau$ diagram in figure 12 of \cite{2008ApJS..179..360G}, except two Z sources, i.e. GX 17+2 and Cyg X-2. As for the Z source GX 17+2, \cite{2009ApJ...699...60L} suggested that the type-\uppercase\expandafter{\romannumeral1} X-ray bursts occurring at the Eddington accretion rate might be due to not only the accretion rate, but also the chemical composition of the accreted matter from its companion star. Moreover, many X-ray bursts were observed when persistent emission was between 0.1 and 0.5 Eddington accretion rate \citep{2012ApJ...748...82L}.

Recently, several authors argued that the accretion emission is synchronously enhanced during X-ray bursts, possibly resulting from the Poynting-Robertdon drag \citep{2013ApJ...772...94W,2015ApJ...801...60W}. However, \cite{2014ApJ...791L..39J} found that in some cases of the hard state, the particles in the accretion disk will be ejected out to infinity, implying the decrease of persistent emission. Such cases are also found for two detected X-ray bursts in this work, i.e. the X-ray bursts \# 12 and 13, as mentioned above. 

Here, we investigate the persistent emission of SAX J1748.9-2021. Simply, we assume that the spectral model of accretion emission during the X-ray bursts is not changed \citep{2016MNRAS.456.4256D}, and the anisotropy factor keeps unvaried. Moreover, we take the NS mass and radius to be 1.4 $\mathrm{M_{\odot}}$ and 10 km, respectively, and then the gravitational redshift 1+z = 1.31. Taking account of the anisotropic emission, the bolometric persistent luminosity $\mathrm{L_{per}}$ can be calculated as follow \citep{1988ApJ...324..995F}
\begin{equation}
\mathrm{L_{per}=4\pi d^2 \xi_{p}F_{per}},
\label{per}
\end{equation}
where $\xi_{\rm p}$ is the anisotropy factor for the persistent emission. Taking the value of $\xi_{\rm p}$ to be $\sim$0.5 \citep{2016ApJ...819...47H} and a distance of 8.5 kpc to the source \citep{1994A&AS..108..653O}, the persistent emission luminosity ($\rm {L_{per}}$) during the X-ray bursts spans a range of $\sim$(0.4-1.9)$\mathrm{\times\,10^{37}\,erg\,s^{-1}}$ or $\sim$(1.0 - 4.9)\%$\mathrm{L_{edd}}$, where $\mathrm{L_{edd}}$ = 3.8$\mathrm{\times\,10^{38}\,erg\,s^{-1}}$. Therefore, the persistent emission of SAX J1748.9-2021 satisfies the accretion condition for the type-\uppercase\expandafter{\romannumeral1} X-ray bursts to occur \citep{2006ApJ...648L.123C}.

\subsection{Burst fuel}

As mentioned above, these X-ray bursts occur at (1.0 - 4.9)\% $\mathrm{\dot{M}_{edd}}$ persistent emission.
In this case, hydrogen is burned into helium steadily through the hot CNO cycle, which means that a pure-helium burst may be ignited by the 3$\alpha$ process \citep{2009A&A...504..501T}.

We now discuss the observed recurrence times of the X-ray bursts.
The data gaps between bursts \#6-7 are minimum, which results in the lowest deviation in consequence.
Thus, we only analyze the observed recurrence times of burst \#6-7. 
Gaps ($\sim$ 2000 s) between bursts \#6-7 indicate that it is possible that X-ray bursts were missed.
Assuming that this is not the case, the observed recurrence time $\tau_{\rm rec}$ of burst \#7 is $\sim$ 2.2 hours.
Burst \#7 exhibits a fast rise ($\sim$ 0.5 s) and a short burst duration ($\sim$ 15 s), which indicates hydrogen-poor material at ignition.
Furthermore, the ratio of the integrated persistent flux to the totally burst energy for burst \#7,
\begin{equation}
\mathrm{\alpha = \tau_{rec} F_{per}/E_{total}},
\end{equation}
can be calculated to be $\sim$ 188.7, which usually indicates a helium-rich X-ray burst.
Then, the observed $\mathrm{Q_{nuc}\,=\,c^{2}\,z/\alpha}$ $\approx$ 1.6 MeV nucleon$^{-1}$ = 1.6+0.4X MeV nucleon$^{-1}$ where c is the speed of light and $\mathrm{X}$ is the mean hydrogen-fraction of the fuel layer.
Then, X $\sim$ 0.
All these phenomena prove that our assumption that X-ray bursts were not missed is reasonable.

\cite{2017ApJ...844...53C} suggested that the companion star of SAX J1748.9-2021 is a main-sequence star.
Thus, we suggest that the hydrogen-fraction of accreted material, $\mathrm{X_{0}}$, is $\sim$ 0.7.
For burst \#7,  hydrogen burns stably as fast as it is accreted, the hydrogen burning time is given by \cite{2016ApJ...819...46L}
\begin{equation}
\mathrm{t_{CNO} = 9.8\,\,(\frac{X_{0}}{0.7})\,\,(\frac{Z_{CNO}}{0.02})^{-1} hr},
\label{eq:delta_t_CNO}
\end{equation}
where $\mathrm{t_{CNO}}$ $\sim$  $\mathrm{\tau_{rec}}$, $\mathrm{Z_{CNO}}$ is the CNO metallicity.
In the following text, we assume that these 13 X-ray bursts originate from the pure-helium ocean ($\mathrm{X}$ = 0) and $\mathrm{Z_{CNO}}$ is kept constant. 

Moreover, we are trying to constrain type-\uppercase\expandafter{\romannumeral1} X-ray burst fuel by comparing the differences between the observed ignition depth and the critical column depth if hydrogen burns stably as fast as it is accreted.
The observed ignition depth is given by 
\begin{equation}
\mathrm{y_{ign} = E_{total} (1+z) (4 \pi R_{NS}^{2} \epsilon_{nuc})^{-1}},
\label{eq:ign-depth}
\end{equation}
where $\mathrm{\epsilon_{nuc}}$ is the nuclear energy released per unit mass that can be calculated from the nuclear energy released per nucleon by $\mathrm{\epsilon_{nuc}}$ = $\mathrm{Q_{nuc}}$ $\mathrm{\times\,10^{18}\,erg\,g^{-1}}$ \citep{2009A&A...504..501T}, the total burst energy $\mathrm{E_{total}}$ can be calculated by
$\mathrm{E_{total} = 4 \pi d^2 \xi_{b} E_{b}}$,
i.e. $\mathrm{E_{total}}$ $\sim$ 0.04 and (0.60-1.12)$\times$10$^{39}$ erg for burst \#10 and type-\uppercase\expandafter{\romannumeral1} X-ray bursts, respectively.
Here $\xi_{\rm b}$ is the anisotropy factor for the burst emission \cite[$\xi_{\rm b}$ $\sim$ 0.5,][]{2016ApJ...819...47H}.
The observed ignition depths $\mathrm{y_{ign}}$ are $\sim$ 0.03 and (0.36 - 0.73) $\mathrm{\times\,10^{8}\,g\,cm^{-2}}$ for burst \#10 and type-\uppercase\expandafter{\romannumeral1} X-ray bursts, respectively.
Then, if hydrogen burns stably as fast as it is accreted, giving a critical column depth \citep{2000ApJ...544..453C} by
\begin{equation}
\mathrm{y_{d} = X_{0} \dot{m}_{per}E_{H}/\epsilon_{H}} 
\end{equation}
where $\mathrm{\epsilon_{H}}$ is $\sim$ 5.8$\mathrm{\times\,10^{15}\,\times\,Z_{\rm CNO}\,erg^{-1}\,s^{-1}}$\citep{1965qssg.conf...17H} and $\mathrm{E_{H}}$ is 6.0 $\mathrm{\times\,10^{18}\,erg\,g^{-1}}$ \citep{1981ApJS...45..389W}
and $\mathrm{\dot{m}_{per}}$ is the local accretion rate per unit area for persistent emission before the X-ray bursts that can be calculated by $\mathrm{\dot{m}_{per} =  L_{per} (1+z) (4\pi R_{NS}^{2} (G M_{NS}/R_{NS}))^{-1}}$, i.e. $\sim$ (2.2-10.5) $\mathrm{\times\,10^{3}\,g\,cm^{-2}\,s^{-1}}$ for all X-ray bursts.

We find that $\mathrm{y_{ign}}$ > $\mathrm{y_{d}}$ for type-\uppercase\expandafter{\romannumeral1} X-ray bursts \#11, 12 and 13, which suggests that hydrogen is completely depleted, leaving only helium for the explosion.
However, $\mathrm{y_{ign}}$ $\le$ $\mathrm{y_{d}}$ for other type-\uppercase\expandafter{\romannumeral1} X-ray bursts, which means these type-\uppercase\expandafter{\romannumeral1} X-ray bursts ignite in a helium-rich environment.
Assuming that the fuel for all type-\uppercase\expandafter{\romannumeral1} X-ray bursts is pure-helium, the properties of each type-\uppercase\expandafter{\romannumeral1} X-ray burst vary little with the accretion rate \citep{2009ApJ...699...60L}.
However, the duration of burst \#12 is obviously shorter than those of other non-PRE X-ray bursts, which suggests that pure-helium is not the case in SAX J1748.9-2021.
Thus, we finally suggest that type-\uppercase\expandafter{\romannumeral1} X-ray bursts in SAX J1748.9-2021 originate from a helium-rich or pure-helium ocean.

\subsection{Irregular X-ray burst}

Among the 13 found X-ray bursts, the X-ray burst \#10 behaves very differently, especially lacking a cooling tail. Such non-cooling X-ray bursts have been observed in four other NS-LMXBs, which show the Z source behaviors (\citetalias{2011ApJ...733L..17L}), while this source, SAX J1748.9-2021, resembles an atoll source somewhat \citep{2009ApJ...690.1856P}. \citetalias{2011ApJ...733L..17L} proposed that it is the ratio of the burst peak flux to the persistent emission flux, i.e. the parameter $\beta$ ($\beta$ $\equiv$ $\mathrm{F_{l}/F_{per}}$), rather than solely $\mathrm{F_{l}}$ or $\mathrm{F_{per}}$, that determine whether cooling tails are present or not. When $\beta$ is less than a threshold, non-cooling tails are observed. The small value of $\beta$ implies the  relatively larger value of $\mathrm{F_{per}}$ and meanwhile the relatively less value of $\mathrm{F_{l}}$. As listed in Table~\ref{tab:table info}, among the 13 X-ray bursts detected in this work, the peak flux of burst \#10 is the least, its persistent emission flux is relatively high, and then $\beta$ shows the least value, i.e. $\sim$1.6$_{-0.8}^{+0.8}$. Therefore, it is reliable that the X-ray burst \#10 could be a non-cooling burst. 

In our analysis, similarly to fit the spectra of those cooling bursts, we use a BB model to fit the time-resolved spectra of this non-cooling burst and the fittings are statistically good, which might indicate that the non-cooling burst is originated from the same mechanism as the cooling bursts, i.e. the thermonuclear burning on the NS surface, as suggested by \citetalias{2011ApJ...733L..17L}. As shown by the top panel of Fig.~\ref{fig:Flux}, the non-cooling burst is found in the episode with the relatively high $\dot{\rm M}$ of the source. \citetalias{2011ApJ...733L..17L} argued that when $\dot{\rm M}$ increases, the NS photosphere will be hotter, which might smear out the burst cooling tail. On the other hand, \citetalias{2011ApJ...733L..17L} suggested that with increasing of $\dot{\rm M}$, the burned mass will be less and thus, the ignition column depth ($\mathrm{y_{ign}}$) will be lower, resulting in colder and less energetic bursts. As listed in Table~\ref{tab:table info}, among the 13 bursts, the burst fluence ($\rm {E_{b}}$) of burst \#10 is the least. With the value of $\rm {E_{b}}$, according to Eq. \ref{eq:ign-depth}, the inferred ignition depth of burst \#10 is $\sim$0.03$\mathrm{\times10^{8}g\,cm^{-2}}$, certainly being the least ignition depth among the13 bursts, because $\mathrm{y_{ign}\,\propto\,E_{total}}$. Therefor, our result somewhat conforms to the mechanism of \citetalias{2011ApJ...733L..17L} responsible for non-cooling bursts.

\section*{Acknowledgements}

We thank the anonymous referee for her or his constructive comments and suggestions, which helped us to improve the presentation of this paper. This research has made use of the data obtained through the High Energy Astrophysics Science Archive Research Center (HEASARC) On-line Service, provided by NASA/Goddard Space Flight Center (GSFC). This work is supported by the National Program on Key Research and Development Project (Grant Nos. 2016YFA0400804, 2016YFA0400803), the Natural Science Foundation of China (Grant Nos. 11673023, 11733009), and the 2014 Project of Xinjiang Uygur Autonomous Region of China for Flexibly Fetching in Upscale Talents.







\label{lastpage}
\end{document}